\documentclass{article}
\usepackage{mlspconf,amsmath,graphicx}
\usepackage{comment}
\usepackage{algorithm,algorithmic}     
\usepackage{caption}
\usepackage{subcaption}
\usepackage{comment}
\usepackage{adjustbox}
\usepackage{amsfonts} 
\usepackage{multirow}
\usepackage{nicematrix}
\usepackage{booktabs}
\usepackage{lscape}
\DeclareMathOperator*{\minimize}{minimize}
\DeclareMathOperator*{\argmin}{argmin}
\copyrightnotice{979-8-3503-7225-0/24/\$31.00 \copyright 2024 IEEE}

\begin{document}



\newcommand{\matr}[1]{\mathbf{#1}}
\title{Discriminative community detection for multiplex networks}
    \name{Meiby Ortiz-Bouza and Selin Aviyente \thanks{This work was supported in part by the National Science Foundation under CCF-2006800 and the Air Force Office
of Scientific Research under FA9550-23-1-0224.}}
    \address{Department of Electrical and Computer Engineering, Michigan State University\\                       ortizbou@msu.edu, aviyente@egr.msu.edu}
    \ninept

    \maketitle
         \begin{abstract}
Multiplex networks have emerged as a promising approach for modeling complex systems, where each layer represents a different mode of interaction among entities of the same type. A core task in analyzing these networks is to identify the community structure for a better understanding of the overall functioning of the network. While different methods have been proposed to detect the community structure of multiplex networks, the majority deal with extracting the consensus community structure across layers. In this paper, we address the community detection problem across two closely related multiplex networks. For example in neuroimaging studies, it is common to have multiple multiplex brain networks where each layer corresponds to an individual and each group to different experimental conditions. In this setting, one may be interested in both learning the community structure representing each experimental condition and the discriminative community structure between two groups. In this paper, we introduce two discriminative community detection algorithms based on spectral clustering. The first approach aims to identify the discriminative subgraph structure between the groups, while the second one learns the discriminative and the consensus community structures, simultaneously. The proposed approaches are evaluated on both simulated and real world multiplex networks.
        \end{abstract}
        

\maketitle


\vspace{-0.1in}
\section{Introduction}\label{sec:introduction}
A multiplex network is a multilayer network where all layers share the same set of nodes with edges representing different interactions \cite{kivela_multilayer_2014}. These multiplex networks model complex systems like living organisms, human societies, and transportation systems \cite{smith2019using}. 

Community detection is a core task in network analysis, where communities are defined as groups of nodes that are more densely connected to each other than they are to the rest of the network \cite{fortunato2016community}. Detecting the community structure is useful for understanding the structure and function of complex networks.  Existing multiplex community detection approaches typically assume that the community structure is the same across layers and find the partition that best fits all layers \cite{magnani2021community}. Thus, they do not differentiate between communities that are common across layers from those that are unique to each layer. In addition, in a lot of applications, one may be interested in comparing the community structure of two multiplex networks. In general, in settings where we have multiplex networks constructed from different conditions, e.g., a treatment and a control experiment, we are interested in visualizing and exploring communities that are specific to one multiplex network. For example, in the study of brain networks through multiplex representations \cite{de2017multilayer}, each layer may correspond to a different subject, and each group may correspond to a different population, e.g., healthy vs. disease. In these settings, it is of great importance to not only understand each group's community structure but to also determine the network components, i.e., communities, that discriminate between the two groups. 

In this paper, we introduce a discriminative community detection approach based on spectral clustering for detecting to identify community structures that distinguish between two multiplex networks. We introduce two different formulations: the first focuses on minimizing the normalized cut of the difference between the two groups with an additional regularization term that ensures that the projection distance between the discriminative subspaces is maximized; the second offers a more comprehensive approach where the consensus, discriminative and individual layerwise subspaces are learned simultaneously across the two groups. These two methods are evaluated on synthetic and real multiplex networks, including functional brain networks, comparing two experimental conditions.
\vspace{-0.1in}
\subsection{Related Work}
The problem of community detection in multiplex networks is closely tied to the literature in multiview clustering \cite{chao2021survey}, which deals with the problem of clustering the data points given multiple sets of features. The main approach to multiview clustering is to optimize the objective function to find the best clustering solution for the given data with $N$ samples and $m$ views, yielding a membership matrix $\matr{H}\in \mathbb{R}^{N\times k}$ that indicates groups membership. Some examples of this approach include multiview spectral clustering \cite{dong2013clustering}, multiview subspace clustering \cite{brbic2018multi}, multiview NMF clustering \cite{liu2013multi} and canonical correlation analysis based methods \cite{chaudhuri2009multi}.  The method proposed here is most similar to multiview spectral clustering, which constructs a similarity matrix and minimizes the normalized cut between clusters. 
However, existing multiview spectral clustering methods focus on learning either consensus or both layer-specific and consensus cluster structures \cite{kumar2011co}. Thus, there is no direct emphasis on differentiating between two groups of multiview data.

Another class of methods that are closely related to the proposed framework are contrastive principal component analysis (cPCA) \cite{abid2017contrastive} and discriminative principal component analysis (dPCA) \cite{chen2018nonlinear}. These methods deal with the dimensionality reduction problem similar to PCA. However, unlike PCA which copes with one dataset at a time, they analyze multiple datasets jointly. They extract the most
discriminative information from one dataset of particular interest, i.e., target data, relative to the other(s), i.e.,  background data. The method proposed in this paper can be thought of as an extension of cPCA and dPCA from the Euclidean domain to the graph domain, where the discriminative subspaces now correspond to the discriminative community structure.
\vspace{-0.1in}
\section{Background}
\vspace{-0.1in}
\subsection{Graph clustering and subspace representation}
Let $\mathcal{G}=(V, E, \matr{A})$ be a graph where $V$ is the node set with $|V|=N$, $E$ is the edge set and $\matr{A}\in \mathbb{R}^{N\times N}$ is the adjacency matrix. In this paper, we use undirected (symmetric) weighted and binary adjacency matrices. For a weighted adjacency matrix, $A(i,j)\in [0,1]$, and for a binary adjacency matrix, $A(i,j)\in\{0,1\}$. A graph can also be represented by its normalized Laplacian matrix $\matr{L}=\matr{D}^{-1/2}(\matr{D-A})\matr{D}^{-1/2}$,
with $\matr{D}$ being the diagonal degree matrix where the diagonal entries are $D_{ii}=\sum_{j} A_{ij}$. 

One of the fundamental tasks in analyzing large-scale networks is community detection, which aims to uncover groups of nodes with higher connectivity amongst themselves compared to the rest of the network. 
Given a graph with adjacency matrix $\matr{A}$,  communities can be detected by minimizing the normalized cut \cite{von2007tutorial}, formulated through spectral clustering as follows \cite{ng2001spectral}:
\begin{equation}
\underset{\matr{U},\matr{U}^\top\matr{U}=\matr{I}}{\minimize} \hspace{0.1in}\text{tr}(\matr{U}^\top \matr{L}\matr{U}),
    \label{eq:SC}
\end{equation}
\noindent where $\matr{U}\in \mathbb{R}^{N\times k}$ is an embedding matrix with $k$ being the number of communities. This is the standard form of a trace minimization problem, and the solution is given by $\matr{U}$ whose columns are the $k$ eigenvectors of $\matr{L}$ corresponding to the smallest eigenvalues.  The graph partition is then found by applying $k$-means to the rows of $\matr{U}$.  Each row of $\matr{U}$ represents the coordinates of the corresponding vertex in a lower-dimensional space, encapsulating connectivity information of the vertices in the original graph.
\vspace{-0.1in}
\subsection{Multiplex Network Community Detection}
Multiplex networks can be represented using a finite sequence of graphs $\{\mathcal{G}_l\}$, where $l \in \{1,2,\ldots, L\}$, $\mathcal{G}_l=(V, E_l, \matr{A}_{l})$ \cite{cozzo2015structure}. $V$ is the set of nodes which is the same for all layers, $E_l$ and $\matr{A}_l\in \mathbb{R}^{n\times n}$ are the edges set and the adjacency matrix for layer $l$, respectively. 
A large group of community detection methods for multiplex networks aim to find a consensus community structure across all layers by first merging the layers and then applying a single-layer community detection algorithm to the aggregated networks. 
When the networks are aggregated through the mean operation, the trace minimization problem in Eq. \eqref{eq:SC} can be written as \cite{kumar2011co}:
\vspace{-0.1in}
\begin{equation}
\underset{\matr{U}^* \in \mathbb{R}^{N\times k},{\matr{U}^*}^\top\matr{U}^*=\matr{I}}{\minimize} \hspace{0.1in}\text{tr}({\matr{U}^*}^\top \sum_{l=1}^L\matr{L}_l\matr{U}^*),
    \label{eq:mxSC}
\end{equation}
\noindent where $\matr{L}_l\in \mathbb{R}^{N\times N}$ is the graph Laplacian matrix for layer $l$. The goal is to find a subspace $\matr{U}^*$ that is representative of all the layers in the multiplex network, and the consensus community structure can be found by applying $k$-means to this $\matr{U}^*$.
\vspace{-0.1in}
\section{Discriminative Community Detection Methods}
\begin{figure*}[h!]
\centering
\begin{minipage}[c]{.3\linewidth}
\raggedleft
     \includegraphics[width=0.85\linewidth]{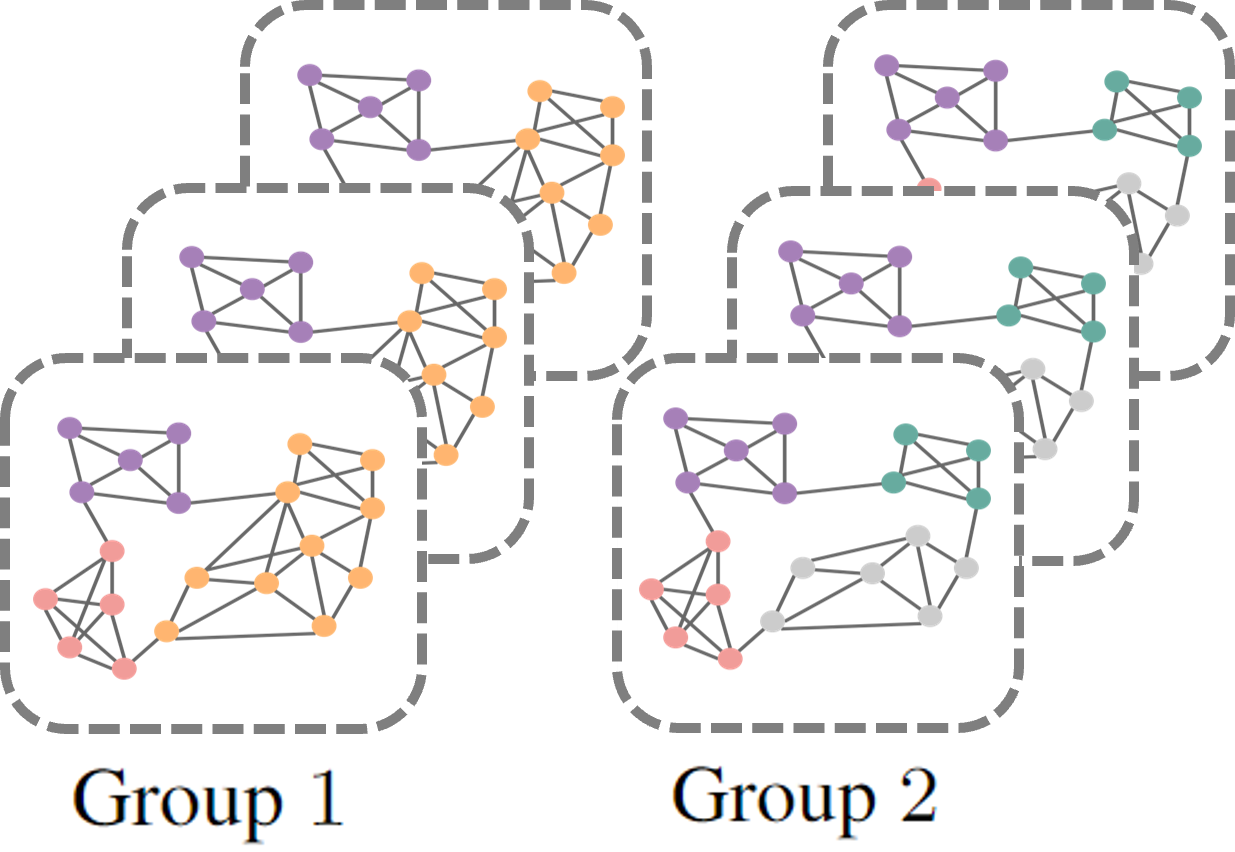}
\end{minipage}
     \begin{minipage}[c]{.12\linewidth}
      \includegraphics[width=0.75\linewidth]{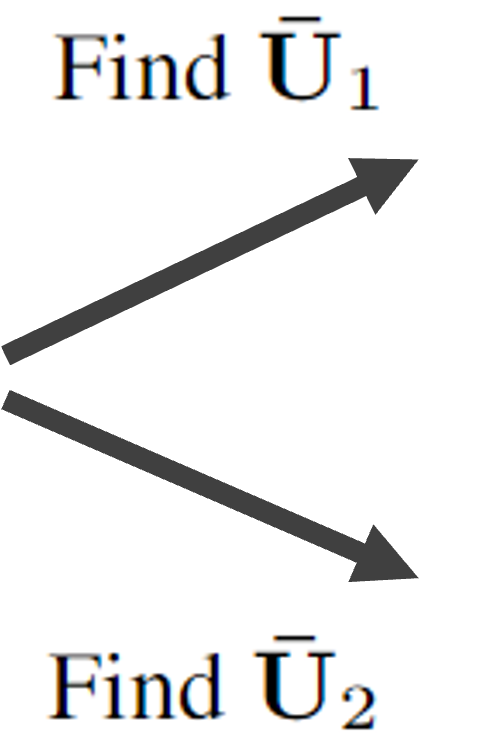}
      \end{minipage}
     \begin{minipage}[c]{.5\linewidth}
    \includegraphics[width=0.95\linewidth]{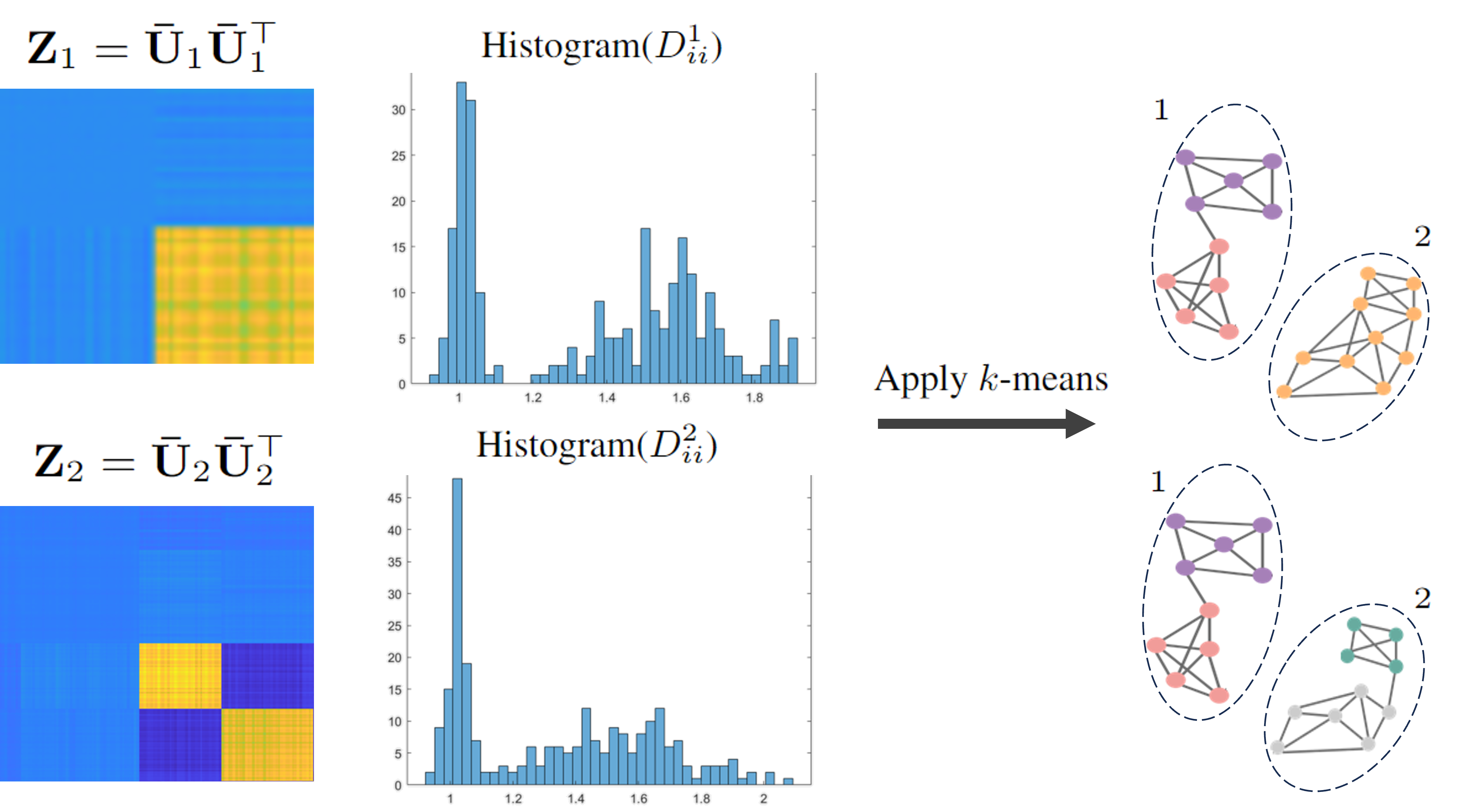}
    \end{minipage}
     \caption{Overview of the proposed methods. Discriminative embedding matrices are learned for each group, $\matr{\bar{U}}_1$ and $\matr{\bar{U}}_2$. $k$-means with $k=2$ is applied to the degrees of $|\matr{Z}_1|$ and $|\matr{Z}_2|$ to separate the nodes in the shared and in the discriminative subspace, respectively.}
    \label{fig:fram}
\end{figure*}
Given two multiplex networks $\mathcal{G}^1_l=(V^1, E^1_l, \matr{A}^1_{l})$ and $\mathcal{G}^2_m=(V^2, E^2_m, \matr{A}^2_{m})$ with  $l \in \{1,2,\ldots, L\}$ and  $m \in \{1,2,\ldots, M\}$ and graph Laplacians $\matr{L}^1_l$ and $\matr{L}^2_m$, the goal is to extract two embedding subspaces, $\matr{\bar{U}}_1$ and $\matr{\bar{U}}_2$, that discriminate between the two multiplex networks with respect to the other. We propose two formulations for achieving this goal.
\vspace{-0.1in}
\subsection{Multiplex Discriminative Spectral Clustering (MX-DSC)}
In the first approach, we focus on obtaining $\matr{\bar{U}}_1$ and $\matr{\bar{U}}_2$ that discriminate between the groups. Let $\matr{\bar{U}}_1 \in \mathbb{R}^{N\times k_{1}}$ be the embedding subspace that minimizes the normalized cut of the first group, i.e., minimize $\text{tr}(\matr{\bar{U}}_{1}^\top (\sum_{l=1}^L\matr{L}_l^1)\matr{\bar{U}}_{1})$, while maximizing the second group's normalized cut, i.e., maximizing $\text{tr}(\matr{\bar{U}}_{1}^\top(\sum_{m=1}^M\matr{L}_m^2)\matr{\bar{U}}_{1})$. These two goals can be simultaneously satisfied through the following optimization:
\vspace{-0.1in}
\begin{equation*}
    \begin{split}
        \underset{\matr{\bar{U}}_1, \matr{\bar{U}}_1^\top \matr{\bar{U}}_1=\matr{I}}{\minimize}\hspace{3mm}\text{tr}(\matr{\bar{U}}_{1}^\top (\sum_{l=1}^L\matr{L}_l^1)\matr{\bar{U}}_{1}) -\alpha\text{tr}(\matr{\bar{U}}_{1}^\top (\sum_{m=1}^M\matr{L}_m^2)\matr{\bar{U}}_{1}),
    \end{split}
\end{equation*}
\noindent where $\matr{\bar{U}}_1$ captures what is discriminative in the first multiplex network with respect to the other. Similarly, we can define $\matr{\bar{U}}_2$ as the embedding matrix that contains information about the discriminative subspace of the second multiplex network with respect to the first.

Considering the two embedding matrices jointly, $\matr{\bar{U}}_1\in \mathbf{R}^{N\times k_1}$ and $\matr{\bar{U}}_2\in \mathbf{R}^{N\times k_2}$, results in  
\vspace{-0.1in}
\begin{equation}
    \begin{split}
        \underset{\matr{\bar{U}}_1\in \mathbb{R}^{N\times k_1}, \matr{\bar{U}}_2\in \mathbb{R}^{N\times k_2}}{\minimize}\hspace{3mm}\text{tr}(\matr{\bar{U}}_{1}^\top (\sum_{l=1}^L\matr{L}_l^1-\alpha\sum_{m=1}^M\matr{L}_m^2)\matr{\bar{U}}_{1})\\
        +\text{tr}(\matr{\bar{U}}_{2}^\top (\sum_{m=1}^M\matr{L}_m^2-\alpha\sum_{l=1}^L\matr{L}_l^1)\matr{\bar{U}}_{2})
        -\gamma\text{tr}(\matr{\bar{U}}_{1}\matr{\bar{U}}_{1}^\top\matr{\bar{U}}_{2}\matr{\bar{U}}_{2}^\top),\\
        \text{s.t. } \matr{\bar{U}}_1^\top \matr{\bar{U}}_1=\matr{I}, \matr{\bar{U}}_2^\top \matr{\bar{U}}_2=\matr{I},
    \end{split}
    \label{eq:fwdisc1}
\end{equation}
\noindent where the first term determines 
$\matr{\bar{U}}_1$ that discriminates the first multiplex network from the second, the second term defines $\matr{\bar{U}}_2$ that discriminates the second network from the first, and the last term is a regularization that maximizes the projection distance between $\matr{\bar{U}}_1$ and $\matr{\bar{U}}_2$. The hyperparameters $\alpha$ and $\gamma$ control the level of discrimination and the dissimilarity between the two subspaces, respectively. 

The optimization problem in \eqref{eq:fwdisc1} can be solved in an alternating manner, first solving for $\matr{\bar{U}}_1$ and then for $\matr{\bar{U}}_2$. 
\begin{equation}
    \begin{split}
   \matr{\bar{U}}_1^{(k+1)}&:=\underset{\matr{\bar{U}}_1\in \mathbb{R}^{N\times k_1}
   }{\argmin} \text{tr}(\matr{\bar{U}}_{1}^\top (\sum_{l=1}^L\matr{L}_l^1-\alpha\sum_{m=1}^M\matr{L}_m^2\\
   &\hspace{0.5in}-\gamma\matr{\bar{U}}_{2}^{(k)}\matr{\bar{U}}_{2}^{{(k)}^\top})\matr{\bar{U}}_{1}),\\
 \matr{\bar{U}}_2^{(k+1)}&:=\underset{\matr{\bar{U}}_2\in \mathbb{R}^{N\times k_2}
}{\argmin} \text{tr}(\matr{\bar{U}}_{2}^\top (\sum_{m=1}^M\matr{L}_m^2-\alpha\sum_{l=1}^L\matr{L}_l^1\\
 &\hspace{0.5in}-\gamma\matr{\bar{U}}_{1}\matr{\bar{U}}_{1}^\top)\matr{\bar{U}}_{2})\\
  &\text{s.t. } \matr{\bar{U}}_1^\top \matr{\bar{U}}_1=\matr{I}, \matr{\bar{U}}_2^\top \matr{\bar{U}}_2=\matr{I}.
    \end{split}
    \label{eq:fwdisc2}
\end{equation}
The solution to updating $\matr{\bar{U}}_1$  is the eigenvectors corresponding to the $k_1$ smallest eigenvalues of $ (\sum_{l=1}^L\matr{L}_l^1-\alpha\sum_{m=1}^M\matr{L}_m^2-\gamma\matr{\bar{U}}_{2}\matr{\bar{U}}_{2}^\top)$, which is the global optimum solution to the $\matr{\bar{U}}_1$ sub-problem in \eqref{eq:fwdisc2} \cite{boyd2004convex}. The solution for $\matr{\bar{U}}_2$ can be found in a similar manner, and it is the global optimum for the $\matr{\bar{U}}_2$ sub-problem in \eqref{eq:fwdisc2}. We solve iteratively for both variables until convergence. 

\vspace{-0.1in}
\subsection{Multiplex Discriminative and Consensus Spectral Clustering (MX-DCSC)}
In this section, we propose a formulation where we learn both the discriminative subspaces between groups, $\matr{\bar{U}}_1$ and $\matr{\bar{U}}_2$, while also learning the consensus subspaces, $\matr{U}_{1}^*\in\mathbf{R}^{N\times k_{t_1}}$ and $\matr{U}_{2}^*\in\mathbf{R}^{N\times k_{t_2}}$ and the individual layerwise embeddings, $\matr{U}_{1_l}\in\mathbf{R}^{N\times k_{t_1}}$ and $\matr{U}_{2_m}\in\mathbf{R}^{N\times k_{t_2}}$, within each group.

For the discriminative part, we propose to use a variation of Eq. \eqref{eq:fwdisc1}, where we find $\matr{\bar{U}}_1$ that captures what is discriminative in the first multiplex network with respect to the other. We can define the squared projection distance between the target representative subspace $\matr{\bar{U}}_1$ of the first group and the
individual subspaces of the second group, $\matr{U}_{2_m}$ as in \cite{dong2013clustering}
\vspace{-0.1in}
\begin{equation*}
    \begin{split}
    d^2_{proj}(\matr{\bar{U}}_1, \{\matr{U}_{2_m}\}_{m=1}^M)&=\sum_{m=1}^M d^2_{proj}(\matr{\bar{U}}_1, \matr{U}_{2_m}) \\
    &=\sum_{m=1}^M(k- \text{tr}(\matr{\matr{\bar{U}}_{1}\bar{U}}_{1}^\top\matr{U}_{2_m}\matr{U}_{2_m}^\top))\\
    &=kM-\text{tr}(\matr{\matr{\bar{U}}_{1}\bar{U}}_{1}^\top\matr{U}_{2_m}\matr{U}_{2_m}^\top).
    \end{split}
\end{equation*}
\indent We want to find a $\matr{\bar{U}}_1$ that minimizes the trace in Eq. \eqref{eq:mxSC} for the graph Laplacians of its group while maximizing its projection distance with the individual subspaces of the second group. Combining these two goals yields the following cost function
\vspace{-0.1in}
\begin{equation*}
    \begin{split}
    \mathcal{L}_{\text{dis}}(\matr{\bar{U}}_1)=\text{tr}(\matr{\bar{U}}_{1}^\top (\sum_{l=1}^L\matr{L}_l^1+\alpha\sum_{m=1}^M\matr{U}_{2_m}\matr{U}_{2_m}^\top)\matr{\bar{U}}_{1}),
    \end{split}
\end{equation*}
\noindent where the regularization parameter $\alpha$ balances the trade-off between the two terms.

To learn the community structure of each layer, we use  trace minimization corresponding to spectral clustering
\begin{equation*}
    \begin{split}
\mathcal{L}_{\text{lw}}(\matr{U}_{1_l}) =\text{tr}(\matr{U}_{1_l}^\top\matr{L}_l^1\matr{U}_{1_l}).
    \end{split}
\end{equation*}

Finally, in order to capture the consensus community structure for each group we use the multiview spectral clustering formulation in \cite{dong2013clustering}
\begin{equation*}
    \begin{split}
\mathcal{L}_{\text{con}}(\matr{U}_{1}^*)=\text{tr}({\matr{U}_{1}^*}^\top (\sum_{l=1}^L\matr{L}_l^1-\beta\sum_{m=1}^M\matr{U}_{1_l}\matr{U}_{1_l}^\top)\matr{U}_{1}^*).
    \end{split}
\end{equation*}

Combining these three terms, $\mathcal{L}_{\text{dis}}$, $\mathcal{L}_{\text{lw}}$ and $\mathcal{L}_{\text{con}}$, for each group and the regularization term that maximizes the projection distance between $\matr{\bar{U}}_1$ and $\matr{\bar{U}}_2$, we propose the following formulation for MX-DCSC, to find $\matr{\bar{U}}_1\in \mathbb{R}^{N\times k_1}$, $ \matr{\bar{U}}_2\in \mathbb{R}^{N\times k_2}$, $\matr{U}_{1_l}\in \mathbb{R}^{N\times k_{t_1}}$, $ \matr{U}_{2_m}\in \mathbb{R}^{N\times k_{t_2}}$, $\matr{U}_{1}^*\in \mathbb{R}^{N\times k_{t_1}}$, and $\matr{U}_{2}^*\in \mathbb{R}^{N\times k_{t_2}}$ 

\begin{equation*}
    \begin{split}
        \underset{\matr{\bar{U}}_1, \matr{\bar{U}}_2,\matr{U}_{1_l}, \matr{U}_{2_m}, \matr{U}_{1}^*, \matr{U}_{2}^*}{\minimize}\hspace{3mm}\text{tr}(\matr{\bar{U}}_{1}^\top (\sum_{l=1}^L\matr{L}_l^1+\alpha\sum_{m=1}^M\matr{U}_{2_m}\matr{U}_{2_m}^\top)\matr{\bar{U}}_{1})\\
        +\text{tr}(\matr{\bar{U}}_{2}^\top (\sum_{m=1}^M\matr{L}_m^2+\alpha\sum_{l=1}^L\matr{U}_{1_l}\matr{U}_{1_l}^\top)\matr{\bar{U}}_{2}) -\gamma\text{tr}(\matr{\bar{U}}_{1}\matr{\bar{U}}_{1}^\top\matr{\bar{U}}_{2}\matr{\bar{U}}_{2}^\top)
        \end{split}
        \end{equation*}
        \begin{equation*}
        \begin{split}
        +\sum_{l=1}^L\text{tr}(\matr{U}_{1_l}^\top\matr{L}_l^1\matr{U}_{1_l})
        +\sum_{m=1}^M\text{tr}(\matr{U}_{2_m}^\top\matr{L}_m^2\matr{U}_{2_m})\\
        \end{split}
        \end{equation*}
        \begin{equation*}
        \begin{split}
        +\text{tr}({\matr{U}_{1}^*}^\top (\sum_{l=1}^L\matr{L}_l^1-\beta\sum_{l=1}^L\matr{U}_{1_l}\matr{U}_{1_l}^\top)\matr{U}_{1}^*)\\
        \end{split}
        \end{equation*}
        \begin{equation}
        \begin{split}
        +\text{tr}({\matr{U}_{2}^*}^\top  (\sum_{m=1}^M\matr{L}_m^2-\beta\sum_{m=1}^M\matr{U}_{2_m}\matr{U}_{2_m}^\top){\matr{U}_{2}^*} )\\
        \text{s.t. } \matr{\bar{U}}_1^\top \matr{\bar{U}}_1=\matr{I}, \matr{\bar{U}}_2^\top \matr{\bar{U}}_2=\matr{I},  \matr{U}_{1_l}^\top \matr{U}_{1_l}=\matr{I}, \matr{U}_{2_m}^\top\matr{U}_{2_m}=\matr{I}\\
        \text{for } l=1,2,\cdots,L \text{ and } m=1,2,\cdots, M.
    \end{split}
    \label{eq:fwdisc3}
\end{equation}

The optimization problem in \eqref{eq:fwdisc3} can be solved in an alternating manner as follows  
\begin{equation*}
\begin{split}
   {\matr{\bar{U}}_1}^{(k+1)}:=&\underset{\matr{\bar{U}}_1
   }{\argmin} \hspace{0.1in} \text{tr}(\matr{\bar{U}}_{1}^\top (\sum_{l=1}^L\matr{L}_l^1+\alpha\sum_{m=1}^M\matr{U}_{2_m}^{(k)}{\matr{U}_{2_m}^{(k)}}^\top\\
   &\hspace{1in}-\gamma\matr{\bar{U}}_{2}^{(k)}\matr{\bar{U}}_{2}^{{(k)}^\top})\matr{\bar{U}}_{1}),\\
 \matr{\bar{U}}_2^{(k+1)}:=&\underset{\matr{\bar{U}}_2
 }{\argmin} \hspace{0.1in} \text{tr}(\matr{\bar{U}}_{2}^\top (\sum_{m=1}^M\matr{L}_m^2+\alpha\sum_{l=1}^L\matr{U}_{1_l}^{(k)}{\matr{U}_{1_l}^{(k)}}^\top\\
 &\hspace{1in}-\gamma\matr{\bar{U}}_{1}^{(k)}\matr{\bar{U}}_{1}^{{(k)}^\top})\matr{\bar{U}}_{2}),
 \end{split}
 \end{equation*}
 \begin{equation*}
 \begin{split}
  \matr{U}_{1_l}^{(k+1)}:=& \underset{\matr{U}_{1_l}
  }{\argmin} \hspace{0.1in} \text{tr}(\matr{U}_{1_l}^\top (\matr{L}_l^1+\alpha\matr{\bar{U}}_{2}^{(k+1)}\matr{\bar{U}}_{2}^{{(k+1)}^\top}\\
  &\hspace{1in}-\beta{\matr{U}_{1}^*}^{(k)}{{\matr{U}_{1}^*}^{(k)}}^\top)\matr{U}_{1_l}),\\
  \matr{U}_{2_m}^{(k+1)}:=& \underset{\matr{U}_{2_m}
  }{\argmin} \hspace{0.1in}  \text{tr}(\matr{U}_{2_m}^\top (\matr{L}_m^2+\alpha\matr{\bar{U}}_{1}^{(k+1)}\matr{\bar{U}}_{1}^{{(k+1)}^\top}\\
     &\hspace{1in}-\beta{\matr{U}_{2}^*}^{(k)}{{\matr{U}_{2}^*}^{(k)}}^\top)\matr{U}_{2_m}),
\end{split}
\end{equation*}
\begin{equation*}
\begin{split}
 \matr{U}_{1}^*{^{(k+1)}}:=&\underset{\matr{U}_{1}^*
 }{\argmin} \hspace{0.1in}  \text{tr}({\matr{U}_{1}^*}^\top (\sum_{l=1}^L\matr{L}_l^1\\
 &\hspace{0.5in}-\beta\sum_{l=1}^L\matr{U}_{1_l}^{(k+1)}\matr{U}_{1_l}^{{(k+1)}^\top})\matr{U}_{1}^*),\\ 
 \matr{U}_{2}^*{^{(k+1)}}:=&\underset{\matr{U}_{2}^*
 }{\argmin} \hspace{0.1in}  \text{tr}({\matr{U}_{2}^*}^\top (\sum_{m=1}^M\matr{L}_m^2\\
&\hspace{0.5in}-\beta\sum_{m=1}^M\matr{U}_{2_m}^{(k+1)}\matr{U}_{2_m}^{{(k+1)}^\top})\matr{U}_{2}^*),
\end{split}
\end{equation*}
\begin{equation}
\begin{split}
 \text{s.t. } \matr{\bar{U}}_1^\top \matr{\bar{U}}_1=\matr{I}, \matr{\bar{U}}_2^\top \matr{\bar{U}}_2=\matr{I},  \matr{U}_{1_l}^\top& \matr{U}_{1_l}=\matr{I}, \matr{U}_{2_m}^\top\matr{U}_{2_m}=\matr{I}\\
       \text{for } l=1,2,\cdots,L \text{ and }& m=1,2,\cdots, M.
    \end{split}
\end{equation}

\subsection{Finding the embedding dimensions}
In most clustering algorithms, the number of communities ($k$) is an input parameter. This is typically addressed by testing different $k$ values and selecting the best based on a performance metric. In this paper, the embedding dimensions, i.e., discriminative and consensus,  are determined following the eigengap rule and hierarchical clustering-based method proposed in \cite{ortiz2024community}. First, the total number of communities for each group, $k_{t_1}$ and $k_{t_2}$, is determined using the eigengap rule, followed by spectral clustering to obtain embedding matrices $\matr{U}1$ and $\matr{U}2$, representing community memberships.
An agglomerative hierarchical clustering algorithm is then applied to $\matr{X}=[\matr{U}_1,\matr{U}_2]$ to determine the number of shared communities ($k_c$) and the discriminative communities per group $k_{1}=k_{t_1}-k_c$ and $k_{2}=k_{t_2}-k_c$, as in \cite{ortiz2024community}. This approach is used for both MX-DSC and MX-DCSC, where MX-DSC only uses $k_{1}$ and $k_{2}$ while MX-DCSC uses $k_{1}$, $k_{2}$, $k_{t_1}$, and $k_{t_2}$.
\vspace{-0.1in}
\subsection{Subgraph Identification}
Once the low-dimensional discriminative subspaces, $\matr{\bar{U}}_1$ and $\matr{\bar{U}}_2$, are learned, we can construct  $N\times N$ matrices, $\matr{Z}_1=\matr{\bar{U}}_1\matr{\bar{U}}_1^\top$ and $\matr{Z}_2=\matr{\bar{U}}_2\matr{\bar{U}}_2^\top$ that capture the discriminative subgraphs for each group, as shown in the toy example in Fig. \ref{fig:fram}. We compute the degrees of nodes in both groups as $D_{1_{i}}=\sum_{j} |Z_{1_{ij}}|$ and $D_{2_{i}}=\sum_{j} |Z_{2_{ij}}|$. As seen in the histograms in Fig. \ref{fig:fram}, 
there are two groups of nodes with different degree distributions. 
These two clusters, discriminative and non-discriminative nodes, in each group, are identified by $k$-means with $k=2$ applied to $\matr{D}_1$ and $\matr{D}_2$, where the cluster with the low degree nodes corresponds to the non-discriminative structure, and the cluster with the high degree nodes correspond to the discriminative subgraph.



\vspace{-0.15in}
\subsection{Computational Complexity}
The computational complexity of the algorithm is mostly due to the eigendecompositions at each iteration. At each iteration, we find the embeddings by computing the eigenvectors corresponding to the $k$ smallest eigenvalues of and $N\times N$ matrix, which has a complexity of  $\mathcal{O}(N^2k)$. Therefore, the total complexity of the algorithm is dominated by $\mathcal{O}(N^2max\{k_1,k_2,k_{t_1},k_{t_2}\})$.

\vspace{-0.09in}
\section{Experiments}
\vspace{-0.09in}
\begin{table*}[h]
\footnotesize
  \centering
  \vspace{-0.2in}
  \caption{Average AUC Values for Synthetic Data}
    \begin{tabular}{l|cc|c|cc|c|cc}
    \hline
 \multicolumn{3}{c|}{Experiment 1} & \multicolumn{3}{c|}{Experiment 2} & \multicolumn{3}{c}{Experiment 3}  \\
  \hline
  $\mu$ & MX-DSC & MX-DCSC & $p_1$ & MX-DSC & MX-DCSC& $k_c$ &  MX-DSC & MX-DCSC\\
  \hline
  0.1 & 1.0000  & 1.0000  & 0.90 & 0.9980 & 0.9978 & 1 & 0.7052 & 0.7115\\
  0.2 & 0.9864  & 0.9801 & 0.80 & 0.9940 & 0.9988   & 2 & 0.9812 & 0.9773 \\
  0.3 & 0.9763 & 0.9761 & 0.70 & 0.9992 & 0.9999 & 3 & 0.9941 & 0.9877\\
  0.4 & 0.9683 &	0.9723 & 0.60 & 0.9956  & 0.9835 & 4 & 0.9802 & 0.9838 \\
  0.5 & 0.9652 & 0.9722 & 0.50 & 0.9821 & 0.9638  & 5 & 0.9949 & 0.9958\\
  0.6 & 0.9398 & 0.9224 & & &  & 6 & 0.9991& 0.9945\\
  0.7 & 0.9167 &  0.8781 &  & & & & \\
  0.8 & 0.8228 & 0.8118 &  & & & \\
    \hline
    \end{tabular}
    \label{tab:AUC}
\end{table*}
\begin{figure*}[h]
\centering
\centering
\vspace{-0.1in}
    \begin{subfigure}[b]{0.3\linewidth}
    \includegraphics[width=0.99\linewidth]{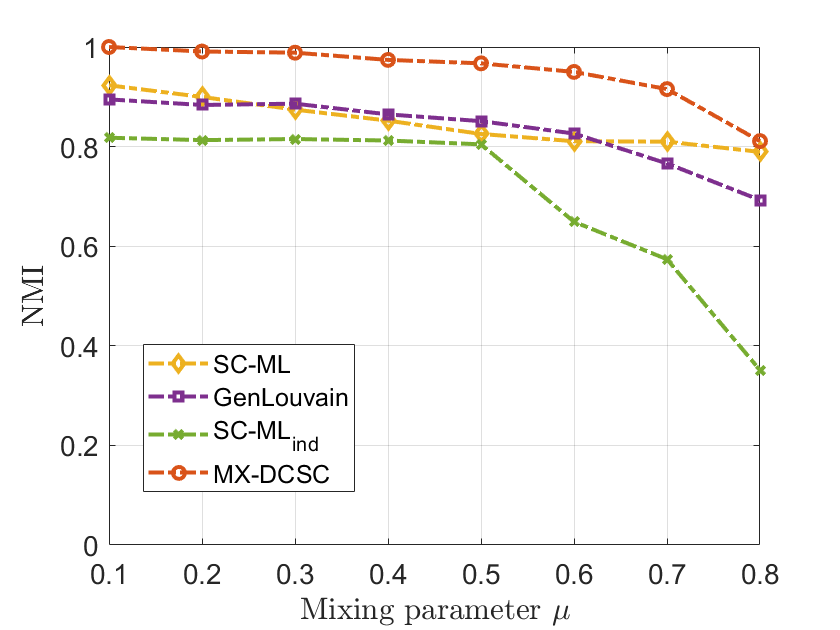}
    \caption{}
    \end{subfigure}
    \centering
    \begin{subfigure}[b]{0.3\linewidth}
    \includegraphics[width=0.99\linewidth]{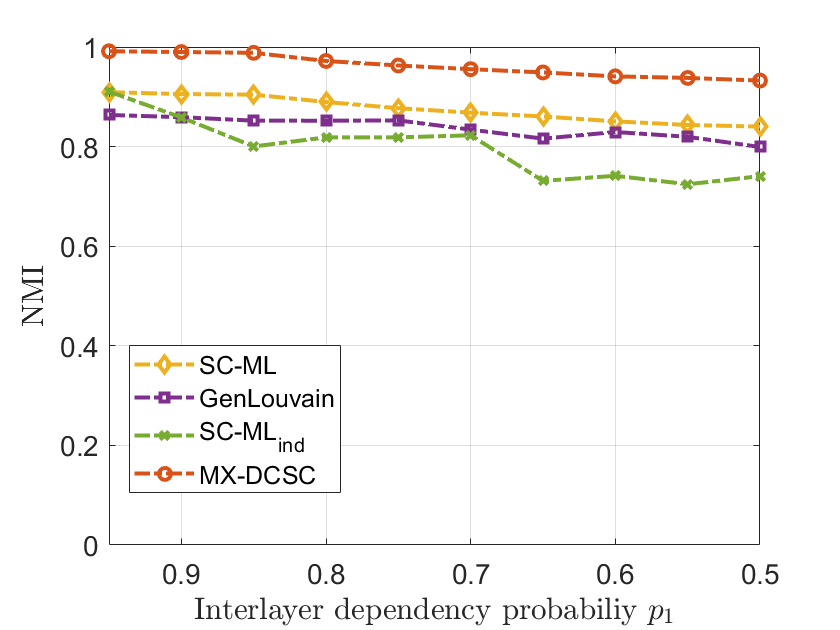}
    \caption{}
    \end{subfigure}
    \centering
    \begin{subfigure}[b]{0.3\linewidth}
    \includegraphics[width=0.99\linewidth]{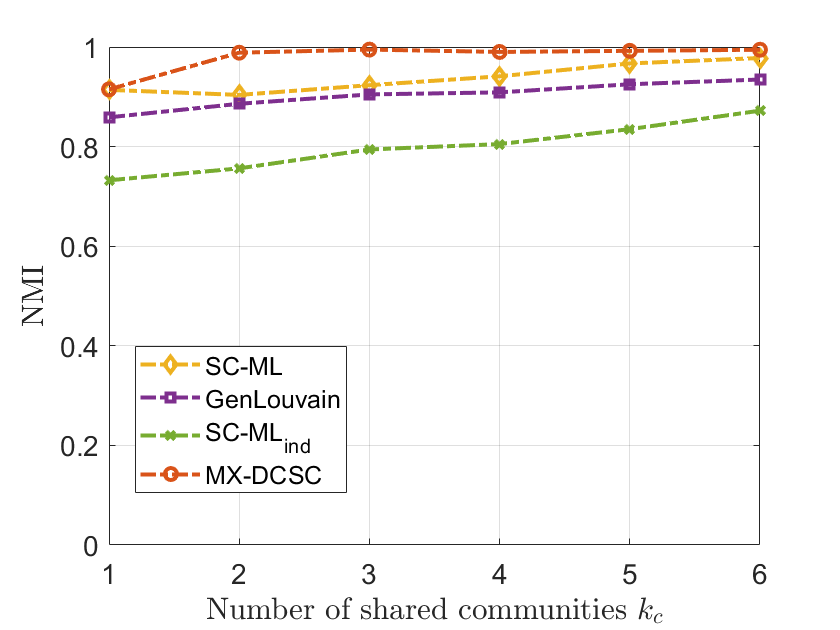}
         \caption{}
    \end{subfigure}
    \vspace{-0.1in}
     \caption{NMI Results for MX-DCSC with respect to other methods: (a) Experiment 1, (b) Experiment 2, (c) Experiment 3 }
    \label{fig:NMIres}
\end{figure*}
\subsection{Synthetic Multiplex Networks}
Multiplex benchmark networks based on the model described in \cite{bazzi2016generative,jeub2016generative} were generated. 
First, a multilayer partition is generated with a user-defined number of nodes, layers, and an inter-layer dependency tensor specifying the layer relationships. Next, for the given multilayer partition, edges in each layer are generated following a degree-corrected block model \cite{karrer2011stochastic} parameterized by the distribution of expected degrees and a community mixing parameter $\mu \in [0,1]$ that controls the network modularity. When $\mu = 0$, all edges lie within communities, whereas $\mu=1$ implies the edges are distributed uniformly.
For multiplex networks, the probabilities in the inter-layer dependency tensor are the same for all pairs of layers and are specified by $p\in [0,1]$. When $p=0$, the partitions are independent across layers while $p=1$ indicates an identical partition across layers.

In this paper, we extend the model described above to generate two multiplex benchmark networks with shared (non-discriminative) and discriminative communities among them two. We first generate the shared communities by randomly selecting $n_c$ nodes across all layers and for both groups and setting the inter-layer dependency probability to $p_1$. 
Next, we independently generate the discriminative communities for each group with the remaining nodes.
\vspace{-0.1in}
\subsubsection{Evaluation}
The performance of MX-DSC is evaluated based on the accuracy of detecting discriminative subgraphs, while MX-DCSC is assessed on both subgraph and community detection accuracy. 
In order to evaluate the performance in detecting the discriminative subgraphs, we use AUC-ROC as the evaluation metric. 
Three experiments with different parameters are repeated 50 times, and the average AUC-ROC for MX-DSC and MX-DCSC are reported in Table \ref{tab:AUC}. We evaluate the accuracy of community detection in terms of Normalized Mutual Information (NMI) \cite{danon2005comparing}. The accuracy of MX-DCSC, which learns the consensus community structure per each group, is compared with existing methods, as shown in Fig. 
\ref{fig:NMIres}. SC-ML \cite{dong2013clustering} is applied, combining both multiplex networks into one, and individually to each multiplex network, SC-ML$_{\text{ind}}$,  and GenLouvain \cite{mucha_community_2010} to the combined networks. 
All the experiments are run with $\alpha,\beta,\gamma \in [0,1]$, and the results with the highest performance are reported.
\vspace{-0.1in}
\subsubsection{Experiment 1: Varying Noise Level, $\mu$}\label{exp:changemu} 
In this experiment, we generated two groups of multiplex networks with $N=256$, 10 layers, and $6$ and $5$ communities per layer in each group, respectively. 
These two multiplex networks have two shared communities, and the number of discriminative communities per group is $k_1=4$ and $k_2=3$, respectively. To evaluate our algorithm's performance under different noise levels, these two networks are generated with varying values of the mixing parameter $\mu\in \{0.1,0.2,0.3,0.4,0.5,0.6,0.7,0.8\}$. The inter-layer dependency probability for the shared communities is $p_1=1$. Table \ref{tab:AUC} shows that both MX-DSC and MX-DCSC have high detection accuracy, with MX-DSC performing slightly better. 
From Fig. \ref{fig:NMIres}, it can be seen that MX-DCSC outperforms existing community detection methods with increasing noise as the consensus community structure takes the discriminative information into account.
\vspace{-0.1in}
\subsubsection{Experiment 2: Change in Variability $p_1$}\label{exp:changep1}
In the second experiment, we evaluated the robustness of the algorithm against variations in the shared community structure by fixing $\mu=0.3$ and varying the inter-layer dependency probability $p_1$, i.e., the shared communities are allowed to vary across groups and layers. Table \ref{tab:AUC} shows that both methods are robust to variations in the shared community structure even when $p_{1}=0.5$, implying that the variations in the shared community structure between the two groups do not affect the discriminative subgraph detection accuracy. Additionally, MX-DCSC performs best in community detection accuracy.
\vspace{-0.15in}
\subsubsection{Experiment 3: Change in  $k_c$}\label{exp:changekc}
In this experiment, we evaluated the robustness of our method to the number of shared communities by fixing $\mu=0.3$ and the total number of communities per layer, while varying $k_c$ from 1 to 6. Both methods are robust to the value of $k_{c}$ except for $k_{c}=1$ since most of the nodes in the two groups will have different community structures, which makes it hard to detect the discriminative subgraphs. In terms of community detection accuracy, MX-DCSC performs well even when the two groups do not share a common community.  
\begin{figure}[h]
\centering
      \includegraphics[width=0.43\linewidth]{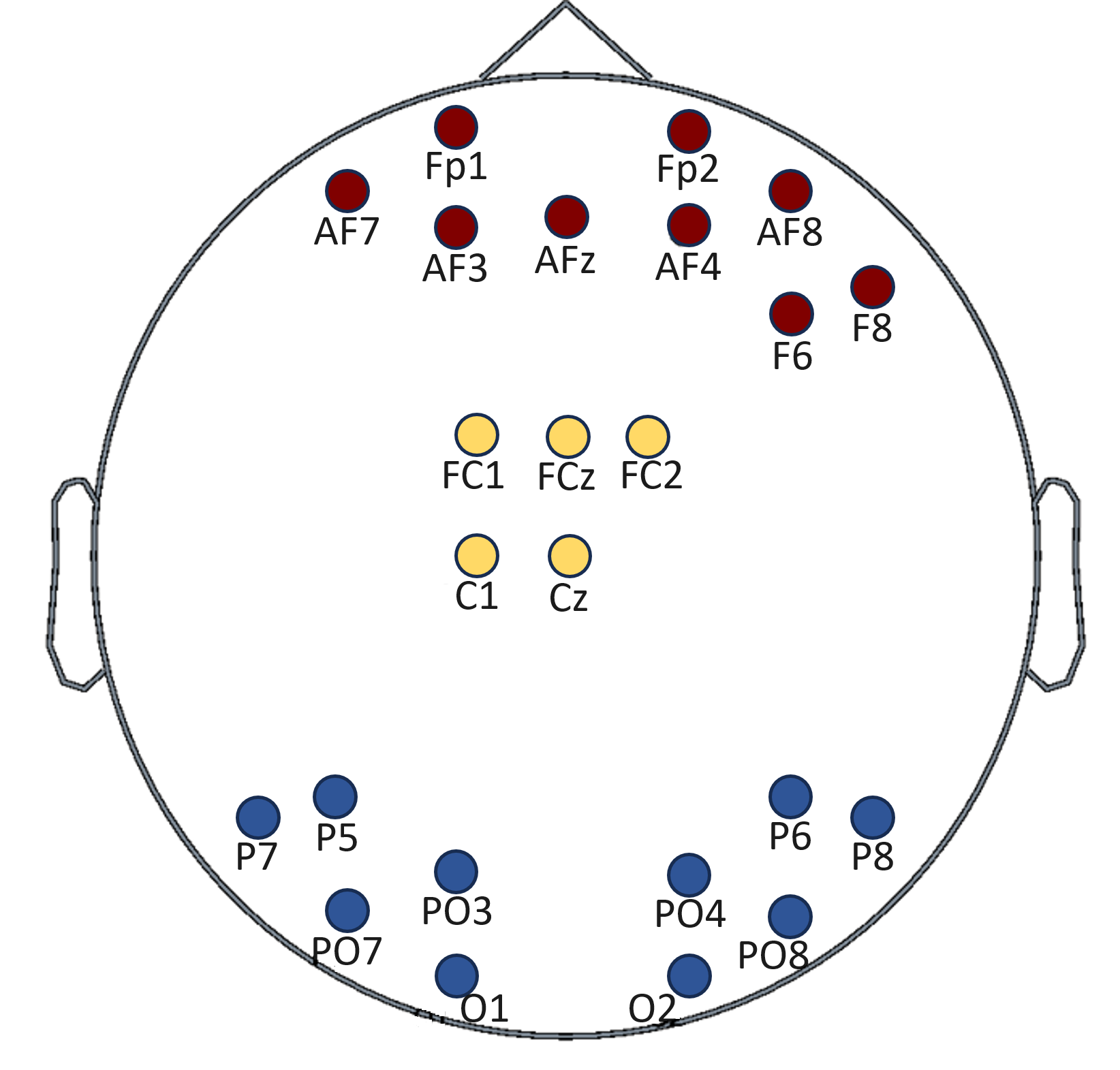}
      \includegraphics[width=0.43\linewidth]{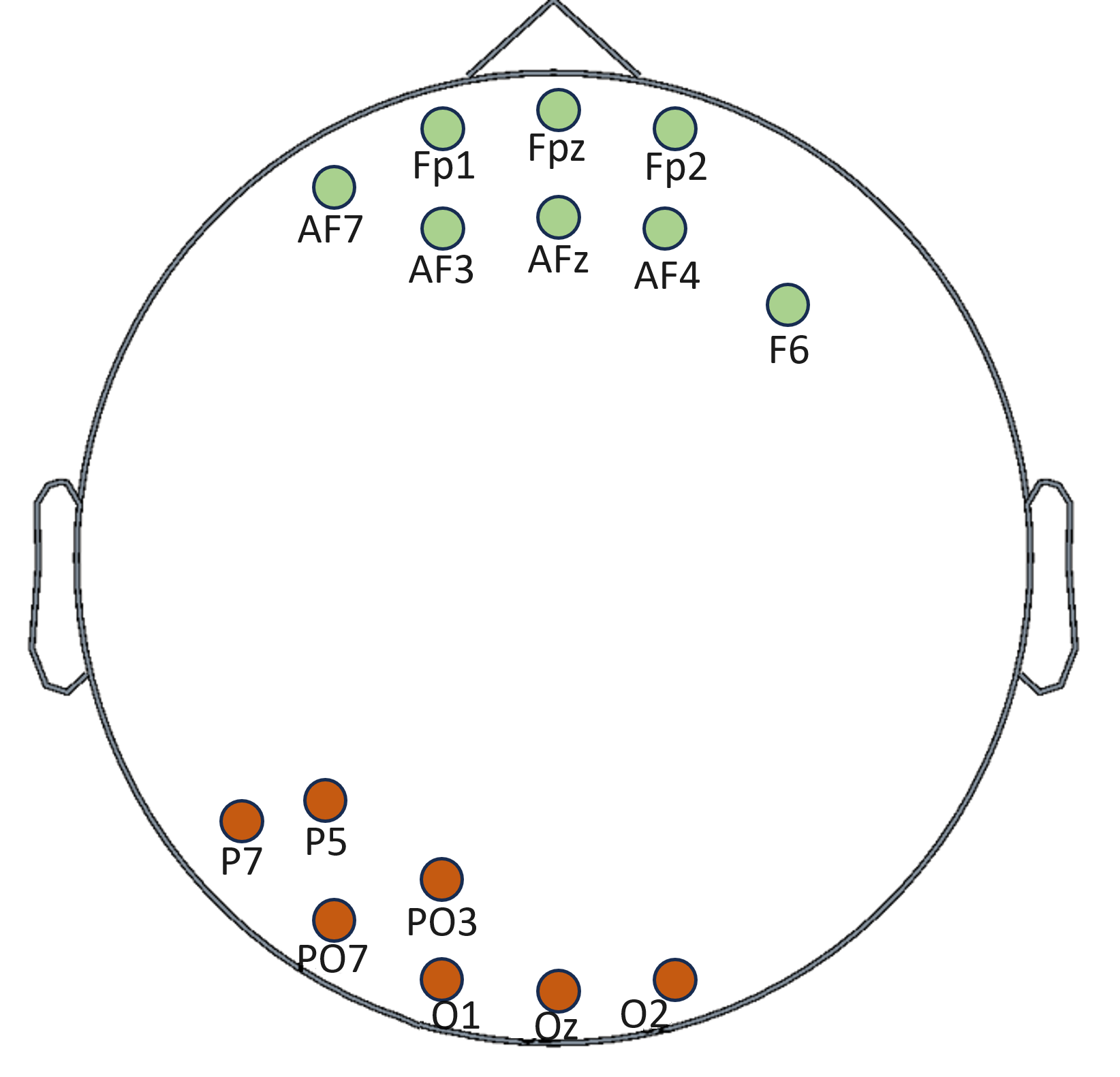}
      \vspace{-0.1in}
     \caption{Discriminative Communities for Error (left) and Correct (right) responses.}
    \label{fig:fram1}
    \vspace{-0.1in}
\end{figure}
\vspace{-0.1in}
\subsection{Electroencephalogram (EEG) Networks}
In this paper, we applied the proposed framework to functional connectivity networks (FCNs) of the brain constructed from EEG data collected from a cognitive control-related error processing study \cite{hall2007externalizing}. Each participant was presented with a string of five letters at each trial. Letters could be congruent (e.g., SSSSS) or 
incongruent stimuli (e.g., SSTSS) and the participants were instructed to 
respond to the center letter with a mouse.  The EEG was recorded following the international 10/20 system for the placement of 64 Ag–AgCl electrodes at a 512 Hz sampling frequency. For each response type (error and correct) the FCNs can be modeled as a multiplex network with 64 nodes (brain regions) and 17 layers (subjects).

MX-DSC is applied with $\alpha=0.5$, $\gamma=0.5$, $k_1=3$, and $k_2=2$  to the multiplex networks corresponding to error and correct responses. Fig. \ref{fig:fram1} shows the discriminative communities corresponding to error and correct responses. For the error response, we detect a discriminative community centered around fronto-central nodes (FCz, FC1, FC2, Cz, C1). This is consistent with prior work showing that medial frontal regions are more activated for
error compared to correct trials \cite{ozdemir2015hierarchical}. The other discriminative communities are similar in both response types and correspond to the parietal-occipital region which is activated due to the visual stimulus. 
\vspace{-0.1in}
\subsection{UCI Handwritten Dataset}
\vspace{-0.05in}
In this section, we evaluate the performance of our method on a multiview dataset, UCI Handwritten Digits\footnote{https://archive.ics.uci.edu/ml/datasets/Multiple+Features}  \cite{Dua:2019}. The UCI Handwritten Digits dataset consists of features of handwritten
digits from (0- 9) extracted from a collection of Dutch utility maps. There are a total of 2000 patterns that have been digitized in binary images, 200 patterns per digit. These digits are represented by six different feature sets: Fourier coefficients of the character shapes, profile correlations, Karhunen-Lo\`{e}ve coefficients, pixel averages in 2 $\times$ 3 windows, Zernike moments, and morphological features. Each layer of the multiplex network represents one of the 6 features. The graphs are constructed using $k$-nearest neighbors graphs with the nearest 50 neighbors and Euclidean distance. For the purposes of this paper, we selected two groups corresponding to digits $1$ and $7$ to construct the first and second multiplex networks, respectively. These are two digits that are often misclustered due to their similar patterns.
\begin{figure*}[h]
\centering
\begin{subfigure}[b]{0.46\linewidth}
\centering
    \includegraphics[width=0.46\linewidth]{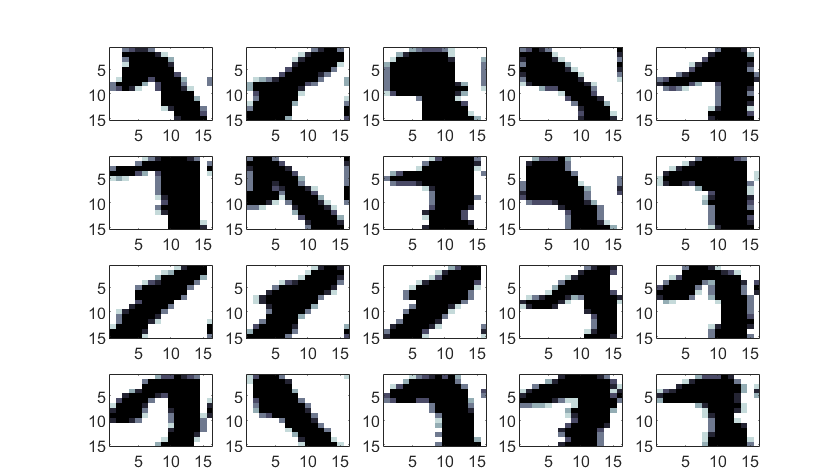}
      \includegraphics[width=0.46\linewidth]{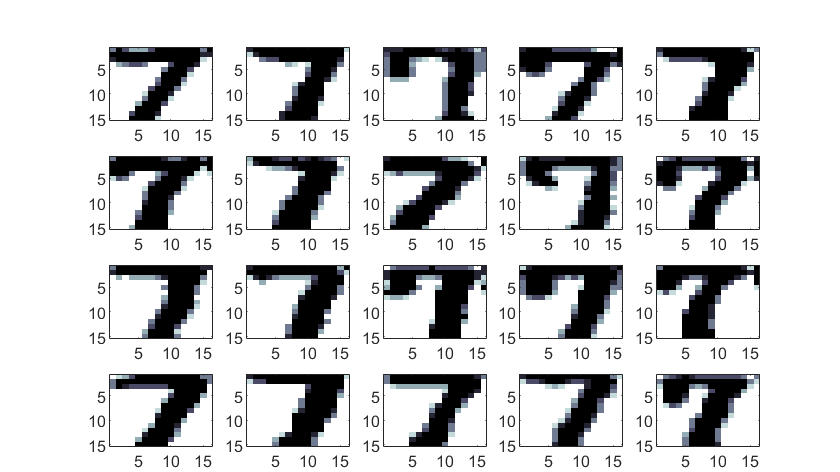}
        \caption{}
    \label{fig:disc17}
    \end{subfigure}
    \begin{subfigure}[b]{0.46\linewidth}
    \centering
    \includegraphics[width=0.46\linewidth]{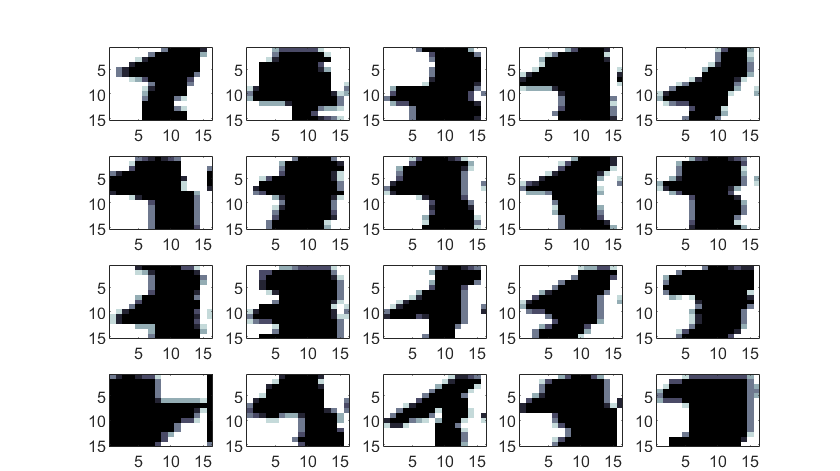}
      \includegraphics[width=0.46\linewidth]{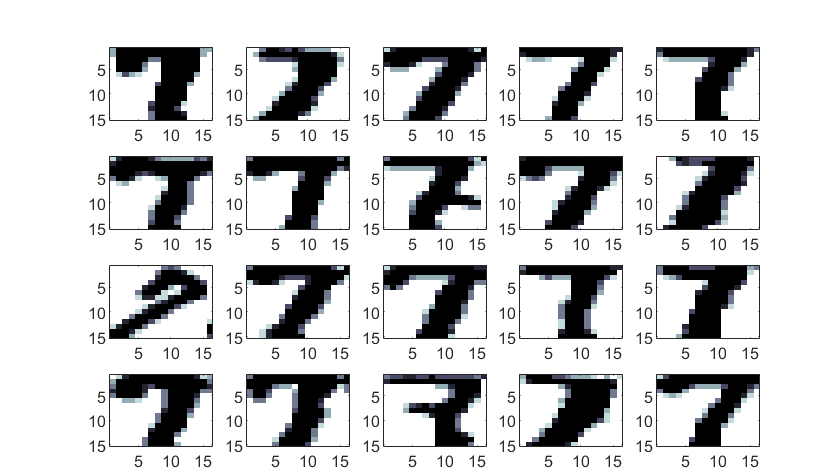}
        \caption{}
    \label{fig:comm17}
    \end{subfigure}
    \vspace{-0.1in}
     \caption{UCI Handwritten dataset results. Images that were selected as (a) discriminative and (b) shared nodes for both groups (digits 1 and 7).}
    \label{fig:digits}
\end{figure*}

MX-DSC is applied to these two multiplex networks each with 6 layers with $\alpha=0.5$, $\gamma=0.5$, $k_1=2$, and $k_2=2$. Fig. \ref{fig:digits} shows the images that were selected as the discriminative and non-discriminative samples between the two groups. The samples in the discriminative subgraph correspond to images where both numbers 1 and 7 are clearly written and well-defined. On the other hand, the samples that are classified as non-discriminative are images with noisy patterns. We also evaluate the performance by applying spectral clustering to the features of the discriminative 1's and 7's and the non-discriminative 1's and 7's separately.  The NMI for the discriminative samples is $0.7037$ whereas it is $0.3091$ for the non-discriminative samples. This shows that our method provides an accurate clustering of samples into discriminative and non-discriminative groups which offers better separability. 


\vspace{-0.1in}
\section{Conclusions}
\vspace{-0.1in}
In this paper, we introduced two spectral clustering based community detection methods for identifying the discriminative subspaces between two multiplex networks. The first method, MX-DSC, focuses on only learning the discriminative subspaces, while the second one, MX-DCSC, learns the discriminative, consensus, and individual layerwise subspaces simultaneously. The evaluation of the methods on simulated data shows that the two methods perform similarly in terms of discriminative subspace detection accuracy. With MX-DCSC, one can also obtain a more accurate community detection performance compared to existing multiplex community detection algorithms. The application of the proposed framework to real data illustrates the possible applications of the method to multivariate network data. Future work will consider the extension of this framework to more than two groups. 
\vspace{-0.1in}
\bibliographystyle{IEEEbib}
\bibliography{refs}
\vspace{-0.1in}
\end{document}